\documentclass[submission,copyright,creativecommons]{eptcs}
\providecommand{\event}{F-IDE 2021} 
\usepackage{underscore}           

\newif\ifREVIEW

\usepackage{paralist}

\usepackage[numbers,sort]{natbib}
\usepackage[inline,draft]{fixme}

\usepackage[utf8]{inputenc}
\usepackage{xcolor}
\usepackage{syntax}
\usepackage{amsmath}
\usepackage{amssymb}
\usepackage{mathpartir}
\usepackage{wrapfig}
\usepackage{braket}
\usepackage{fontawesome}
\usepackage{todonotes}
\usepackage{appendix}
\usepackage{uppaal}
\usepackage{multicol}

\usepackage[numbers]{natbib}
\usepackage{hyperref}

\usepackage{tikz}
\usepackage{fontawesome}
\usetikzlibrary{positioning, arrows,automata,calc,patterns,shapes,shadows}

\usepackage{afterpage}
\usepackage{pgfplots}
\usetikzlibrary{positioning, shapes,patterns,shadows.blur, arrows,decorations.text,arrows,automata,shadows,patterns,chains,bending,arrows.meta}

\definecolor{tfg}{HTML}{EEEEEE}
\definecolor{tbg}{HTML}{333333}
\tikzset{%
	initial text={},
	unit/.style={draw=tfg,line width=0.15pt,rectangle, rounded corners=3pt,line width=0.15pt, color=tbg,fill=tfg!0,font=\tiny,text width=1.5cm, minimum height=0.75cm, align=center, anchor=center,rounded corners=1pt},
}

\newcommand{\cpachecker}{\textsc{CPAChecker}}
\newcommand{\cbmc}{\textsc{CBMC}}
\newcommand{\divine}{\textsc{divine}}
\newcommand{\tapaal}{\textsc{Tapaal}}

\newcommand{\spin}{\textsc{Spin}}
\newcommand{\llvm}{\textsc{LLVM}}

\colorlet{keywords}{green!40!black}
\colorlet{identifiers}{blue}
\colorlet{comments}{purple!40!black}
\colorlet{typecol}{green!40!black}
\newcommand{\registers}{\ensuremath{\mathtt{R}}}
\newcommand{\lang}{\ensuremath{\mathcal{L}}}
\newcommand{\ir}{\textsc{UL}}
\newcommand{\irName}{Uppaal LLVM}
\newcommand{\regi}[1]{\ensuremath{{\color{identifiers}\mathbf{#1}}}}
\newcommand{\cmdname}[1]{\ensuremath{{\color{keywords}\mathbf{#1}}}}
\newcommand{\types}{\ensuremath{\mathbb{T}}}
\newcommand{\intty}[1]{\ensuremath{\color{typecol}\mathbf{i#1}}}
\newcommand{\boolty}{\ensuremath{\color{typecol}{\mathbf{Bool}}}}
\newcommand{\addrty}{\ensuremath{\color{typecol}{\mathbf{Addr}}}}
\newcommand{\assign}{\ensuremath{\leftarrow}}

\newcommand{\bitvec}[1][]{\ensuremath{\mathtt{b}_{#1}}}

\newcommand{\bitone}[1]{\ensuremath{\bar{#1}}}
\newcommand{\bittwo}[1]{\ensuremath{\vec{#1}}}

\newcommand{\onebit}[2]{\ensuremath{[#1]^{\intty{#2}}}}
\newcommand{\twobit}[2]{\ensuremath{[#1]_2^{\intty{#2}}}}

\newcommand{\bitand}{\ensuremath{\&}}
\newcommand{\bitor}{\ensuremath{|}}
\newcommand{\bitlshl}{\ensuremath{<<}}
\newcommand{\bitlshr}{\ensuremath{>>_l}}
\newcommand{\bitashr}{\ensuremath{>>_a}}
\newcommand{\bitconcat}{\ensuremath{\circ}}
\newcommand{\bitadd}{\ensuremath{+}}
\newcommand{\bitsub}{\ensuremath{-}}
\newcommand{\bitdiv}{\ensuremath{/}}
\newcommand{\bitsdiv}{\ensuremath{/_2}}
\newcommand{\bitmult}{\ensuremath{*}}

\newcommand{\prule}[1]{\ensuremath{\mathtt{#1}}}
\newcommand{\typemap}{\ensuremath{\Gamma}}

\newcommand{\memory}{\ensuremath{\delta}}
\newcommand{\mems}{\ensuremath{\Delta}}
\newcommand{\timing}{\ensuremath{\Omega}}

\newcommand{\dom}{\ensuremath{\mathtt{dom}}}

\newcommand{\locs}{\ensuremath{\mathtt{L}}}
\newcommand{\edges}{\ensuremath{\mathtt{E}}}
\newcommand{\initlocation}{\ensuremath{\mathtt{l}}}

\newcommand{\errorstate}{\ensuremath{\dagger}}

\newcommand{\cout}[1]{\ensuremath{c!}}
\newcommand{\cin}[1]{\ensuremath{c?}}

\newcommand{\ta}[1][]{\ensuremath{\mathcal{A}^{#1}}}
\newcommand{\Cplusplus}{C\raisebox{0.5ex}{\tiny\textbf{++}}}

\newcounter{remarkcounter}
\newcommand{\remarkhack}{\smallskip\noindent\refstepcounter{remarkcounter}\emph{Remark \arabic{remarkcounter}.}}

\usepackage{stackengine}

 \ifREVIEW
	\usepackage{lineno}
	\linenumbers
	\newif\ifANONYMOUS
\fi

\lstdefinestyle{customc}{
  belowcaptionskip=0pt,
  breaklines=true,
  frame=none,
  xleftmargin=1em,
  xrightmargin=-0.5cm,
  numbersep=2pt,
 language=c,
  showstringspaces=false,
  basicstyle=\scriptsize\ttfamily,
  keywordstyle=\bfseries\color{keywords},
  commentstyle=\itshape\color{comments},
  identifierstyle=\color{identifiers},
  stringstyle=\color{orange},
  numberstyle=\tiny\color{gray},
  numbers=left, 
}

\title{Analysis of Source Code Using \uppaal{}}
\author{Mitja~Kulczynski
\institute{Kiel University, Kiel, Germany}
\email{mku@informatik.uni-kiel.de}
\and
Axel~Legay
\institute{Univeristy of Louvain, Louvain-la-Neuve, Belgium}
\email{axel.legay@uclouvain.be}
\and
Dirk~Nowotka
\institute{Kiel University, Kiel, Germany}
\email{dn@informatik.uni-kiel.de}
\and
Danny~Bøgsted~Poulsen
\institute{Aalborg University, Aalborg, Denmark}
\email{dannybpoulsen@cs.aau.dk}
}
\def\titlerunning{Analysis of Source Code Using \uppaal{}}
\def\authorrunning{M.~Kulczynski, A.~Legay, D.~Nowotka \& D.B.~Poulsen}
\usepackage{paralist}

\begin{document}

\AtBeginEnvironment{grammar}{\footnotesize}

\maketitle
\begin{abstract}
  In recent years there has been a considerable effort in optimising
formal methods for application to code. This has been driven by tools
such as \cpachecker, \divine, and \cbmc. At the same time tools such
as \uppaal{} have been massively expanding the realm of more traditional
model checking technologies to include strategy synthesis algorithms ---
an aspect becoming more and more needed as software becomes
increasingly parallel. Instead of reimplementing the advances made by
\uppaal\ in this area, we suggest in this paper to develop a bridge
between the source code and the engine of \uppaal. Our
approach uses the widespread intermediate language LLVM and makes
recent advances of the \uppaal{} ecosystem readily available to
analysis of source code.   
\end{abstract}

\section{Introduction}
Over 30 years of research in applying formal methods to program
verification has resulted in a plethora of
tools~\citep{DBLP:conf/cav/BeyerK11,DBLP:conf/tacas/KroeningT14,spin97,BBK+17,DBLP:conf/osdi/CadarDE08,DBLP:journals/sttt/LarsenPY97,DBLP:conf/tacas/DavidJJJMS12}
each being developed 
by different groups.  We could write an entire paper just about the
differences of all these tools, but overall they can  (very) roughly be divided into 
two categories based on their input format:
\begin{inparaenum}
\item tools that accept real source code, and 
\item tools that use their own format based on formal models
  (e.g. Finite Automata, Timed Automata and Petri Nets). 
\end{inparaenum}
In the former category we find tools such as
\cpachecker~\cite{DBLP:conf/cav/BeyerK11},
\cbmc~\citep{DBLP:conf/tacas/KroeningT14},  \divine~\citep{BBK+17} and JavaPathfinder~\citep{DBLP:journals/ase/VisserHBPL03}
focused on locating \emph{programming errors}
while in the latter we find \uppaal~\citep{DBLP:journals/sttt/LarsenPY97},
\tapaal~\citep{DBLP:conf/tacas/DavidJJJMS12}, \textsc{Alloy}~\citep{jackson2019alloy}, \textsc{TLA+}~\cite{chaudhuri2010verifying}
\textsc{Prism}~\citep{Kwiatkowska2011,DBLP:conf/qest/KwiatkowskaNP04}
and \spin~\citep{spin97} focused on finding errors in the
\emph{design} of a system. The tools in each of these categories are
successful in their own 
right, but there is only little flow between the groups. Two notable
exception is \divine~\cite{BBK+17}
that started its life as a general-purpose model
checker and now focused on verifying \llvm\ and   
Zaks and Joshi~\cite{DBLP:conf/spin/ZaksJ08} utilising \spin\ to verify \llvm\
programs.   

In this paper we present our initial work to bridge the gap from
automata-based \uppaal{} models to 
 source-code analysis. We do this by interfacing \llvm-programs with
\uppaal\ using \uppaal{}s extendibility through dynamic link
libraries~\cite{DBLP:conf/iceccs/JensenLLN17,DBLP:conf/birthday/CassezAJ17}. In the dynamic link library resides an
interpreter communicating with \uppaal\  in regards to what
happens with the discrete state-space while \uppaal\ manages the
exploration algorithms and timed aspects of the state space. In this way, we allow reusing the very
efficient state exploration algorithms already implemented in
\uppaal but avoid mapping the entire expressivity of \llvm\ into \uppaal. Furthermore, since an  entire suite of tools is built
around the \uppaal\ core we hope to leverage these tools
in the future. We especially have high hopes for doing  schedulability
analysis of concurrent programs in the future, and for using the
statistical model checking (SMC) engine of 
\uppaal{} to not only speed up the search, but
also as a strategy for finding programming errors in a similar way as
Chockler et al.~\cite{10.1145/2480362.2480588} used SMC for the \emph{satisfiability
  problem}. The usage of simulation-based techniques  should also
alleviate the scalability issue that hampered previous attempts  at reusing  model checking tools for
software verification  techniques.  

Another potential advantage of integrating source code analysis into  \uppaal\
is that environmental behaviours  are easily modelled with the  
stochastic hybrid automata formalism used by \uppaal. Therefore we can
easily change the analysis of a program source code in shifting
environments by changing model parameters.


\section{Program Model}
In our work we are concerned with mapping \llvm\ code to \uppaal, but
we implement the integration to \uppaal\ using our
own intermediate representation called \ir{} (\irName). The main reason for this is two-fold:
firstly the \llvm\ language is huge and trying to cover it entirely is
beyond the scope of this paper, and secondly by basing our translation
on our own intermediate representation makes the translation independent of
the input formalism, and we can --- in principle --- perform analysis for
any input format with \uppaal. From a maintenance point of view it
also makes sense to define the analysis on your own internal
representation as it makes the analysis independent of the input
format: if we used \llvm\ directly we (potentially) have to modify large parts of
our infrastructure for new releases of \llvm\ whereas we by having our
own \ir{} ``just'' need to modify our \llvm\ loading mechanism. 

\medskip
\noindent\emph{Types} In \ir\ we have four different integer types
(\intty{8},\intty{16},{\intty{32} and \intty{64}). Like in \llvm\
integer types are finite width bit vectors with no interpretation
in regards to signedness. Instead each instruction of \ir\ decides
whether it interprets the bit pattern as being 2s-complement encoded
signed number or an unsigned binary number. In addition to integer types \ir\ has a
Boolean type (\boolty )  and an address type (\addrty) that are pointers to places
in memory. We let $\types$  be the set of all types in \ir.

\medskip
\noindent\emph{Instructions} Given a finite set of variables \registers\ we 
denote all instruction sequences of \ir{} by \lang(\prule{Sequence}). All possible instruction
sequences of \ir\ are generated by the EBNF in \autoref{fig:ebnf}. \lang(\prule{Sequence}) 
refers to the language generated by the production rule $\langle\prule{Sequence}\rangle$. A typical instruction sequence are, for example,
\begin{inparaenum}[]
\item an arithmetic expression, 
\item a Boolean comparison operation,
\item or a memory operation.
\end{inparaenum}
Our instructions do not have associated types. A Type in \ir\ is instead associated directly to registers
through a map $\typemap :  \registers \rightarrow \types$. Given this map we can
straightforwardly create a type system, which we omit in this paper.

\medskip
\noindent\emph{Memory}
\ir\ uses a zero-indexed byte-oriented memory layout,  so formally the
memory is just a function $\memory : \mathbb{N} \rightarrow
\mathbb{B}^{8}$. We refer to the set of all possible memory states by
$\mems$. We can update the $n^{\mathrm{th}}$ byte in memory \memory{} by simply modifying the image of $n$ in \memory{}.
Thus $\memory[n \mapsto \bitvec]$ sets the $n^{\mathrm{th}}$  byte to $\bitvec \in\mathbb{B}^{8}$.  
\newcommand{\aalt}{\;\llap{\textbar}\;}
\begin{figure}[tb]
  \centering
  \begin{grammar}
    <\prule{Sequence}> ::= <Internal> <InstrSeq> \aalt <InstrSeq>
    
    <\prule{InstrSeq}> ::= <InstrSeq> <Instr> \aalt <Instr> \aalt <Assigns>

    <\prule{Instr}> ::= <Arith> \aalt <Cast> \aalt <Cmp> \aalt <Memory>
    
    <\prule{Internal}> ::= \cmdname{Assume} \regi{r} \aalt \cmdname{NegAssume} \regi{r} \aalt \cmdname{Assert} 

    <\prule{Assigns}> ::= \regi{r} \aalt \regi{r}~\assign~\cmdname{NonDet} \aalt \regi{r}~\assign \cmdname{Copy} \prule{Op}
    
    <\prule{Arith}> ::= \regi{r}~\assign~\cmdname{Add}~<Op>,~<Op>
    \aalt~\regi{r}~\assign~\cmdname{Sub}~<Op>,~<Op>
    \aalt~\regi{r}~\assign~\cmdname{Div}~<Op>,~<Op>
    \aalt~\regi{r}~\assign~\cmdname{SDiv}~<Op>,~<Op>
    \aalt~\regi{r}~\assign~\cmdname{Mult}~<Op>,~<Op>
    \aalt~\regi{r}~\assign~\cmdname{LShl}~<Op>,~<Op>
    \aalt~\regi{r}~\assign~\cmdname{AShr}~<Op>,~<Op>
    \aalt~\regi{r}~\assign~\cmdname{LShr}~<Op>,~<Op>
  
    <\prule{Cast}> ::= \regi{r}~\assign~\cmdname{SExt}~<Op>
    \aalt~\regi{r}~\assign~\cmdname{ZExt}~<Op>
    \aalt~\regi{r}~\assign~\cmdname{Trunc}~<Op>
    \aalt~\regi{r}~\assign~\cmdname{BoolSExt}~<Op>
    \aalt~\regi{r}~\assign~\cmdname{BoolZExt}~<Op>
    
    <\prule{Cmp}> ::= \regi{r}~\assign~\cmdname{LEq}~<Op>,~<Op>
    \aalt~\regi{r}~\assign~\cmdname{SLEq}~<Op>,~<Op>
    \aalt~\regi{r}~\assign~\cmdname{NEq}~<Op>,~<Op>
    \aalt~\regi{r}~\assign~\cmdname{Eq}~<Op>,~<Op>
    \aalt~\regi{r}~\assign~\cmdname{GEq}~<Op>,~<Op>
    \aalt~\regi{r}~\assign~\cmdname{SGEq}~<Op>,~<Op>
    
    <\prule{Memory}> ::= \regi{r}~\assign~\cmdname{Load}~<Op>
    \aalt~\cmdname{Store} <Op>, <Op>
    
    <\prule{Op}> ::= \regi{r} | $[n]^{\intty{b}}_{2}$ | $[n]^{\intty{b}}$
\end{grammar}
\caption{EBNF for generating instruction sequences of \ir. Let $\regi{r}\in\registers$, $n\in\mathbb{N}$ and $z\in\mathbb{Z}$. The notation ($[n]^{\intty{b}}_{2}$) $[n]^{\intty{b}}$ is our notation encoding a number $n$ into (2s-complement) bit-vector. }
\label{fig:ebnf}
\end{figure}

\medskip
\noindent\emph{Timing Information} A classical Control Flow Automaton (CFA) is a tuple
  $(\registers,\typemap,\locs,\initlocation, \edges)$ where $\registers$ is a set of
  registers, $\typemap : \registers \rightarrow \types$ maps registers
  to types, $\locs$ is a set of control locations, $\initlocation
  \in\locs$ is the initial location and $\edges \subseteq \locs
  \times\lang(\prule{Sequence})\times \locs$ is the set of flow edges
  annotated with instruction sequence to be executed while moving
  along that edge. Such a model is sufficient if we are only concerned with the
``flow'' of a program, but we know  that timing is an just as important
aspect of a program: an airbag has to deploy at the right time, and
not just at some point after a crash. In  a security context it is also
fairly well-known that measuring timing of a program can sometimes
reveal confidential information about it. Even using very low capacity covert timing channels (2 bits per minute)
leaking secret data e.g. a credit card number can be done in less than 30 minutes~\cite{agat2000transforming}. 
For these reasons we want to extend our model with
timing information. To this end assume there exists a function 
$\timing : \lang(\prule{Sequence}) \rightarrow
\mathcal{I}$ --- where $\mathcal{I}$ is the set of intervals in
$\mathbb{R}$ --- that assigns upper and lower bound on the
execution time of an instruction sequence. Notice that for defining
this function it suffices to define the intervals for individual
instructions as the execution time of a sequence is the sum of
the individual instructions. We use the $\timing$ function to enrich
a classical CFA coping with time.

\remarkhack{}
  Expert-knowledge of the execution platform is needed to properly
  asses the execution time of individual instructions. We are not
  concerned with assessing that for now, but do acknowledge this as
  non-trivial and an aspect worth investigating in the future. 
\medskip

\noindent\emph{Transition Semantics}
The first thing to define is the
domain of the types in \ir. \autoref{tab:domains} shows the map-

\begin{wrapfigure}[8]{R}{.3\linewidth}
  \vspace{-0.2cm}
  \centering
  \begin{tabular}{|l|l|}
    \hline 
    Type & Domain \\
    \hline 
    \intty{b}& $\mathbb{B}^b$ \\
    \boolty & $\{\mathtt{true},\mathtt{false}\}$ \\
    \addrty & ${\mathbb{B}^{64}}$ \\
    \hline
  \end{tabular}
  \caption{Semantic domains of types in \ir}\label{tab:domains}
\end{wrapfigure}

\noindent pings in
\ir.  The state that we execute an instruction sequence in
consists of the sequence of instructions and an environment giving values to
the registers $\mathtt{Env} : \registers \rightarrow
\cup_{t\in\types}\dom(t)$. A special state $\errorstate$ signifies an error happened during execution.  Externally the
memory $\memory\in\mems$ and the types of register $\typemap$ are passed on to the transitions.
The result is a modified environment
$\mathtt{Env}'$ and an updated memory $\memory'\in\mems$. Thus the
transition rules take the form $\memory,\typemap \vdash \langle
\mathtt{Inst}, \mathtt{Env} \rangle \rightarrow
\mathtt{Env}',\memory'$.

The transition rules
are obmitted do to space constraints and will not be shown here. They essentially look up values of their
operands, perform the
associated operations in the bit vector logic and assigns the left hand
side to the result. The instructions $\cmdname{SExt}$, $\cmdname{ZExt}$ resp.\,$\cmdname{BoolSExt}$, $\cmdname{BoolZExt}$ and $\cmdname{Trunc}$ sign-extend or zero-extend or truncates the right hand operand to
the type of the left hand side.

\remarkhack{}
  \ir\ does not allow function calls. This is
  because we assume all function calls has been inlined. Inlining
  functions for verification purposes is a common
  practice~\citep{DBLP:conf/tacas/KroeningT14,DBLP:conf/tacas/FalkeMS13}
  and drastically simplifies the interpreter code.  

\smallskip

\noindent An actual transition is performed on a network of CFAs, that is a structure $\,\mathcal{C}_1\; \|\;\dots\;\|\;
\mathcal{C}_n$ where each $\mathcal{C}_i =
(\registers_i,\typemap_i,\locs_,\initlocation^i, \edges_i)$ is a
CFA. Again we assume the existence of a function 
$\timing$. The state of such a network is a tuple
$(s_1,s_2,\dots,s_n,\memory)$ where each $s_i = (l_i,\mathtt{Env}_i,x_i)$, 
$l_i\in\locs$, $\mathtt{Env}_i : \registers_i\rightarrow
\cup_{t\in\types}\dom(t)$, $x_i\in\mathbb{R}$ and $\memory\in\mems$ is
a memory state. Enriching the CFA with a timed behaviour results in specifying the following two transitions:
a state $S = (s_1,s_2,\dots,s_i,\dots,s_n,\memory)$
\begin{enumerate}
\item  can transit to state $S' = (s_1',s_2',\dots,s_i',s_n',\memory')$  (written $S \rightarrow S'$) where $s_i =
(l_i,\mathtt{Env},x_i)$ and $s_i' = (l_i',\mathtt{Env'},0)$ if there exists an edge $e=(l_i,Inst,l_i')\in\edges_i $,  
$\memory,\typemap_i \vdash \langle Inst, \mathtt{Env} \rangle \rightarrow \mathtt{Env}',\memory'$ and $x_i \in\timing(Inst)$.

\item can delay $d$
  time units to state $S' = (s_1',s_2',\dots,s_i',\dots,s_n',\memory)$ 
  (written $S\xrightarrow{d} S'$) if, for all $i$, $s_i =
  (l_i,\mathtt{Env},x_i)$ such that $s_i' = (l_i,\mathtt{Env},x_i+d)$ and there exists an edge $e = (l_i,Inst,l_i')$ such that $x_i+d\in\timing(Inst)$.
\end{enumerate}

\remarkhack{}
  The semantics for networks of CFAs is the traditional interleaving
  semantics. Our semantics, however, 
  considers the execution of each edge atomically and does not
  properly reflect all possible interleavings of a real program ---
  unless each edge has exactly one instruction as in the semantics of
  \llvm\ by Legay et al.~\cite{legay2020automatic}. This is not
  a problem as long we ensure all interleavings of memory accesses are possible. We
  can guarantee this by splitting edges so that an edge  has exactly one
  memory access which is guaranteed to be the last instruction.  

\section{Integration with \uppaal}
\newcommand{\cfaEx}{\resizebox{1cm}{!}{\begin{tikzpicture}	[->,>=stealth',shorten >=1pt,node distance=3cm, semithick]	
                       \tikzstyle{every state}=[ align=center,top color=white, bottom color=black!10, draw=white!10!black!40,drop shadow]
						\tikzset{every node/.style={rectangle,draw=black!10,fill=black!10}}
						\node[state] (A)                        {$1$};
						\node[state,double] (B)[below of=A]    {$2$};
						\node[state] (C)[diamond,below of=B]    {$3$};
						\node[state] (D)[below right of=C]    {$4$};
						\node[state] (E)[below left of=C]    {$5$};
						\node[state] (F)[right of=B]    {$6$};
						\node[state] (G)[right of=F, node distance=3.5cm]    {$7$};

						\path 	(A) edge[->] node  {$x' = 5;$} (B)
                                   (B) edge node  {$x \le 0$}  (F)
                                         edge node  {$x > 0$} (C)
									(C) edge node  {$x > 5$}  (D)
                                         edge node  {$x \le 5$}  (E)
									(D) edge[bend right] node  {$x = x - 2$}  (B)
									(E) edge[bend left]  node  {$x = x - 1$}  (B)
									(F) edge node  {return $0$}  (G);
	\end{tikzpicture}}}

In this section we will describe the general translation of a network
of CFAs $\mathcal{C}_1\; \| \; \dots \; \| \; \mathcal{C}_n$ into a \uppaal\

\begin{wrapfigure}[15]{R}{.45\linewidth}
  \centering
  \vspace*{-0.2cm}
\resizebox{.45\textwidth}{!}{
\begin{tikzpicture}[->,>=stealth',shorten >=1pt,auto,node distance=1cm,
semithick]

\tikzset{every edge/.append style={font=\tiny,color=tbg}}

\draw[tbg,rounded corners=3pt,fill=tbg!1] (-2.05,0.55)  rectangle (2.05,-3.05);
\node[unit]   (0)  at (-1,0)    {\faFileCodeO{}\;LLVM code}; 
\node[unit]   (1)  at (1,0)  {\faMapMarker{}\;Entry point(s)};

\node[unit]   (2)  at (0,-1.25)  {\faFileCodeO{}\;C$^\mathrm{++}$-file};
\node[unit]   (3)  at (3.2,-2.5)  {\faFileO{}\;\texttt{minimc.so}};

\node[unit]   (4)  at (-1,-2.5)  {\faFileTextO{}\;XML specifications};
\node[unit]   (5)  at (1,-2.5)  {\faFileO{}\;\texttt{code.so}};

\node[unit,fill=tbg!3,font=\scriptsize]  (6) at (0,-4) {\faIndustry{}\;\uppaal{}};
\node[font=\tiny]   (7)  at (3.05,-4)  {$\varphi$\;Verification query};

\path (0) edge [out=270,in=180] node [] {}   (2)
      (1) edge [out=270,in=0] node [] {}   (2)
      (2) edge [out=250,in=90] node [] {}   (4)
          edge [out=290,in=90] node [] {}   (5)
      (3) edge [out=180,in=0] node [] {}   (5)
      (4) edge [out=270,in=120] node [] {}   (6)
      (5) edge [out=270,in=70] node [] {}   (6)
      (7) edge [out=180,in=0] node [] {}   (6);
\end{tikzpicture}}
\caption{tool chain}\label{fig:toolchain}
\end{wrapfigure}
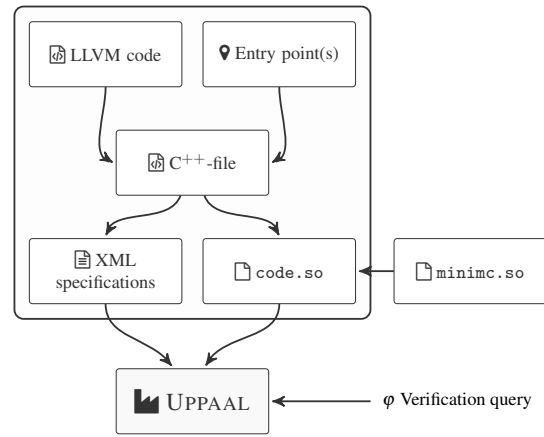

\noindent
Timed Automaton~\citep{DBLP:journals/sttt/DavidLLMP15}. For starters each CFA $\mathcal{C}_i$ is  
represented by a single Timed Automaton \ta[i] with the same graph
structure and stores register values in an integer array
\uppExpr{lState}. It also has one clock \uppExpr{x}. Memory is similarly represented using an integer
array \uppExpr{memory}. In regards to memory there is one last
component we need, and that is an integer stored in the \uppExpr{memory}
locating the next unused byte. For simplification our semantics did not include global
registers, but we are nevertheless including them in our actual
tool. These are stored in a \uppaal\ integer array \uppExpr{glob}.

\remarkhack{}
  On the surface this translation framework might seem a bit
  unnecessarily complicated but it does provide advantages. Firstly,
  since the state of CFAs are kept inside \uppaal\ we can take
  advantage of \uppaal{}s capabilities: namely model checking and
  statistical model checking. Secondly, since \uppaal\ knows about the
  graph structure of the CFA, we can use those locations as part of
  verification queries.

\medskip
\noindent\emph{Automated tool chain}
We developed a tool chain\footnote{\url{https://gitlab.com/dannybpoulsen/uppllvm}}
 that automates building the  
\uppaal\ model along with connecting it to the CFA interpreter (see 
\autoref{fig:toolchain}). 
\noindent The tool accepts an
\llvm\ input-file along with information
about the entry-points of the program (in case of a parallel program
several entry points can be specified). This information is embedded into a \Cplusplus-file which is
compiled and linked against the library
(\texttt{minimc}\footnote{\url{https://github.com/dannybpoulsen/minimc}})
resulting in  a dynamic link library (\texttt{code.so}) 
providing an interface to the interpreter (used in the resulting
\uppaal\ model as discussed),  and also query functions in
regards to the structure of the CFA. 
These latter query functions are 
used by a python script that creates the \uppaal\ XML-file.
Finally the \texttt{code.so} library and XML-file are passed to \uppaal\ in which
verification questions can be posed.
The \texttt{minimc}
library is needed as this converts \llvm\ to our \ir\ structure and
also performs needed syntactical modifications to the CFA like expanding
\cmdname{NonDet} instructions into several edges (one for each
possible value),
ensuring each edge only has one memory instruction,
inline all \llvm\ functions,
identify assert statements and add a special
\texttt{AssertViolation} location that is reached only when an
assert is violated, and add a \texttt{Term} location indicating normal termination of a CFA.

\section{The tool chain in action}
This section is devoted to demonstrating our tool chain. Next to verifying a password validation form which leaks information over time, we analyse a classical mutual exclusion protocol: Petersons Algorithm. This diverse set of examples demonstrates the capabilities of the presented approach.

\subsection{On timing leaks}
Timing is an important aspect of many programs and a major reason for integrating software models like \llvm\ into \uppaal. In this example we identify potential timing leaks in a wrongly implemented password
validation program given in \autoref{lst:timing}. After setting the password to \texttt{abcdef} the program enters a
\begin{wrapfigure}[11]{r}{5cm}
  \centering
  \vspace*{-0.3cm}
  \resizebox{!}{3.9cm}{
  \begin{tikzpicture}
    \begin{axis}[axis line style={draw=none}, xtick pos=left, ytick pos=left, ymajorgrids=true, legend style={draw=none,fill=white},ymin=0, ymax=350,xlabel=Runtime (ms),ylabel=Count of executions]
      \addplot+[ybar interval,mark=no,black!40,fill=black!40] plot coordinates {
        (103.872,1) (104.638498,2) (105.404996,6) (106.171494,3)
        (106.937992,4) (107.70448999999999,5) (108.470988,4)
        (109.237486,13) (110.003984,8) (110.770482,20) (111.53698,13)
        (112.303478,20) (113.069976,30) (113.836474,30)
        (114.602972,41) (115.36947,45) (116.135968,54) (116.902466,68)
        (117.668964,59) (118.435462,87) (119.20196,103)
        (119.968458,107) (120.734956,102) (121.501454,125)
        (122.26795200000001,118) (123.03444999999999,157)
        (123.800948,143) (124.567446,173) (125.333944,184)
        (126.100442,199) (126.86694,226) (127.633438,218)
        (128.399936,250) (129.166434,246) (129.932932,274)
        (130.69943,285) (131.465928,316) (132.232426,286)
        (132.998924,308) (133.765422,313) (134.53192,308)
        (135.298418,297) (136.064916,288) (136.831414,307)
        (137.597912,271) (138.36441,309) (139.130908,272)
        (139.897406,275) (140.663904,294) (141.43040200000002,283)
        (142.1969,249) (142.96339799999998,207) (143.729896,214)
        (144.496394,181) (145.262892,200) (146.02939,162)
        (146.795888,157) (147.562386,146) (148.32888400000002,108)
        (149.095382,109) (149.86187999999999,100) (150.628378,76)
        (151.394876,82) (152.161374,66) (152.927872,67) (153.69437,51)
        (154.460868,41) (155.22736600000002,41) (155.993864,33)
        (156.760362,26) (157.52686,20) (158.293358,19) (159.059856,22)
        (159.826354,13) (160.592852,7) (161.35935,8) (162.125848,6)
        (162.892346,3) (163.658844,9) (164.425342,0) (165.19184,3)
        (165.958338,2) (166.724836,6) (167.491334,1) (168.257832,2)
        (169.02433000000002,1) (169.790828,3) (170.557326,1)
        (171.323824,0) (172.09032200000001,2) (172.85682,1)
        (173.62331799999998,2) (174.389816,1) (175.156314,0)
        (175.92281200000002,0) (176.68931,0) (177.455808,0)
        (178.222306,0) (178.98880400000002,1)   };
    \end{axis}
  \end{tikzpicture}}
  \caption{Runtime distribution of \autoref{lst:timing}.
    }
  \label{fig:distributionTim}
\end{wrapfigure}
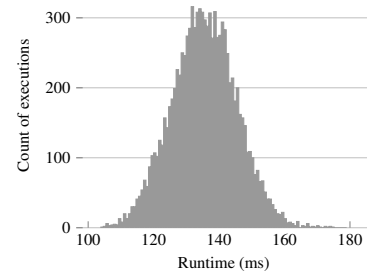
loop reading characters by the function \texttt{read()}. The characters are immediately compared to the expected password \texttt{abcdef}. However, \texttt{read()} is an
undefined function and therefore our toolchain replaces it by a \cmdname{NonDet}
instruction which forces \uppaal{} to search all possible
values of \lstinline{k} in each iteration of the loop. Adding a clock \uppExpr{G} that is never reset to the \uppaal\
model allows us to estimate the run time of all paths through the program
by the \uppaal{} query \begin{center}\scriptsize\texttt{E[<=500;1000] (max: G*(1-main.AssertViolation || main.main_Term))}.\end{center} We went up 500ms within 1000 runs. to \autoref{fig:distributionTim} shows a
distribution plot of the run times revealing a variation in the run times. 
A symbolic analysis (in \uppaal) reveals that a properly terminating program has a
run time in the range  $[125,175]$ ms.  Any execution-time outside
this  range could leak information to an attacker, as it has lower run-time because we broke out of
the loop due to mismatching characters.

\begin{figure}[h]
\begin{multicols}{2}
\begin{lstlisting}[style=customc]%,caption={Program depending on timing.},label=lst:timing,captionpos=b]
void assert (int);
char read ();

#define N 6

int main () {
  char sec[N];
  for (int i = 0; i <N; i++) {
    sec[i] = 'a'+i;
  }
  for (int i = 0; i< N; i++) {
    char k = read();
    if (k != sec[i])
      assert(0);
  }
  return 0;
}\end{lstlisting}
\end{multicols}
\caption{Program depending on timing.\label{lst:timing}}
\end{figure}

\subsection{Verifiying a mutual exclusion protocol}
As a proof of concept we have  implemented a mutual exclusion protocol (Petersons Algorithm) in C  --- see
\autoref{lst:petersons} for an excerpt ---  and ran it through our 
toolchain after first compiling it to \llvm\ using
\texttt{clang}. Using \uppaal\ we are able to verify that the
algorithm  
behaves correctly (i.e. guarantees mutual exclusion).  We have also made an
implementation of Petersons Algorithm  containing an
error  breaking the mutual exclusion property. The error is
initialising the \texttt{*opt.mflag} variable wrongly on line 20. Our
toolchain does correctly find a path showing the mutual exclusion
property is broken in this case. In order to locate the error we had
to first find the CFA location (let us call it \uppLoc{Crit} correspodning to line 45 and 26
respectively, and since \uppaal\ has knowledge about those we can
simply ask \uppaal\ \texttt{E<> (petersons1.Crit \&\& petersons2.Crit)}.

\begin{figure}[h]
\begin{multicols}{2}
\begin{lstlisting}[style=customc]%,caption={Excerpt of Petersons algorithm in C.},label=lst:petersons,captionpos=b]
typedef struct {
  int *mflag;
  int *oflag;
  int *turn;
}Options;

int turn = 0;
int oneflag;
int secondflag;

int crit1 = 0;
int crit2 = 0;

void petersons1 () {  
  Options opt;
  opt.mflag = &oneflag;
  opt.oflag = &secondflag;
  opt.turn = &turn;
  
  *opt.mflag = 1;
  *opt.turn = 1;
  
  while (*opt.oflag  
          && *opt.turn == 1) {
   /* busy wait */
  }
  // critical section
  crit1 = 1;
  // end of critical section
  crit1 = 0;
  *opt.mflag = 0;
}

void petersons2 () {
  Options opt;
  opt.mflag = &secondflag;
  opt.oflag = &oneflag;
  opt.turn = &turn;
  
  *opt.mflag = 1;
  *opt.turn = 0;
  
  while (*opt.oflag  
          && *opt.turn == 0) {
   /* busy wait */
  }
  // critical section
  crit2 = 1;
  // end of critical section
  crit2 = 0;
  *opt.mflag = 0;  
}
\end{lstlisting}
\end{multicols}
\caption{Excerpt of Petersons algorithm in C.\label{lst:petersons}}
\end{figure}
\section{Conclusion}
In this paper we presented our preliminary work towards utilising
\uppaal\  for model checking of \llvm-code. Our current strategy is to
``outsource'' \llvm\ to an external interpreter and allow \uppaal\
to act as a simulation controller. Our initial experiments show the
integration is functional and warrants further investigations. In the future we will investigate mixing our \llvm-verification model with (stochastic) models of the environment. This would allow analysing the
source in different environmental contexts by simply changing
parameters of the model. A use-case could be analysing whether a cruise
controller for a car responds fast enough when moved to
quicker accelerating car which names an important capability of the presented ideas.

\bibliographystyle{eptcs}
\bibliography{bibliography}

\newpage
\end{document}